\documentclass[12pt]{iopart}
\usepackage{iopams}
\begin{document}

\title[Peakon equations with cubic nonlinearity]{Integrable peakon equations with cubic nonlinearity}
\author{Andrew N.W. Hone and Jing Ping Wang\dag} 
\address{\dag\ Institute of Mathematics, Statistics \&
Actuarial Science,
University of Kent,
Canterbury CT2~7NF, UK}
  
\ead{anwh@kent.ac.uk jw83@kent.ac.uk}

\def\underset#1#2{\mathrel{\mathop{#2}\limits_{#1}}}   
\newcommand{\haf}{{\hat{f}}}   
\newcommand{\beq}{\begin{equation}}  
\newcommand{\eeq}{\end{equation}}  
\newcommand{\bea}{\begin{eqnarray}}  
\newcommand{\eea}{\end{eqnarray}}  
\newcommand\la{{\lambda}}  
\newcommand\ka{{\kappa}}  
\newcommand\al{{\alpha}}  
\newcommand\be{{\beta}}  
\newcommand\de{{\delta}}
\newcommand\si{{\sigma}}  
\newcommand\lax{{\bf L}}  
\newcommand\mma{{\bf M}}   
\newcommand\ctop{{\mathcal{T}}}   
\newcommand\hop{{\mathcal{H}}}   
\newcommand\B{{\mathcal B}}  
\newtheorem{Pro}{Proposition}  
\newtheorem{Thm}{Theorem}

\newcommand\ga {{\gamma}}
\newcommand{\rd}{{\mathrm{d}}}  
\newcommand{\sgn}{{\mathrm{sgn}}}  
\def \ring {{\cal R}}

\begin{abstract}   
We present a new integrable partial differential equation found by Vladimir Novikov.
Like the Camassa-Holm and Degasperis-Procesi equations, this new equation 
admits peaked soliton (peakon) solutions, but it has nonlinear terms that are 
cubic, rather than quadratic. 
We give a matrix Lax pair for V. Novikov's equation, 
and show how it is related by a reciprocal transformation to a negative flow in the 
Sawada-Kotera hierarchy. Infinitely many conserved quantities  are found, as well as
a bi-Hamiltonian 
structure. The latter is used to obtain the Hamiltonian form of the finite-dimensional 
system for the interaction of $N$ peakons, and the two-body dynamics ($N=2$) 
is explicitly integrated. Finally, all of this is compared with some 
analogous results for another cubic peakon derived by Zhijun Qiao.  
\end{abstract}   
  
\submitto{J. Phys. A: Math. Gen.}   
  
\maketitle  

\section{Introduction} 

The subject of this paper is the partial differential equation  (PDE)  
\beq\label{full} 
u_t-u_{xxt}+4u^2u_x=3uu_xu_{xx}+u^2u_{xxx}, 
\eeq 
which was discovered very recently by Vladimir Novikov in a 
symmetry classification of nonlocal PDEs with cubic nonlinearity 
\cite{volodya}. 
The perturbative symmetry approach \cite{miknov} yields 
necessary conditions for a PDE to admit infinitely many 
symmetries. Using this approach, Novikov was able to 
isolate the equation (\ref{full}) and find its first few symmetries, 
and he subsequently found a scalar Lax pair for it, proving 
that the equation is integrable.    
Due to the $u_{xxt}$ term on the left hand side 
of (\ref{full}), this equation is not an evolutionary PDE 
for $u$. However, taking the convolution with the 
Green's function $g(x)=\exp (-|x|)/2$ for the Helmholtz operator 
$(1-\partial_x^2)$ gives the nonlocal (integro-differential) 
equation 
$ 
u_t + u^2 u_x +g* [3uu_xu_{xx}+2(u_x)^3+3u^2u_x]=0$.
It is convenient to define a new dependent variable $m$ to be 
the Helmholtz operator acting on $u$, in which case the  
equation (\ref{full}) can be more concisely written as 
\beq \label{vn} 
m_t+u^2m_x+3uu_xm=0, \qquad m=u-u_{xx}. 
\eeq 
Henceforth we work with the above form of the equation. 

The work of Camassa and Holm \cite{ch}, who derived  
the equation 
\beq \label{ch} 
m_t+um_x+2u_xm=0, \qquad m=u-u_{xx} 
\eeq 
from an asymptotic approximation to the Hamiltonian for the 
Green-Naghdi equations in shallow water theory, has 
attracted a lot of interest in the past fifteen years,  
for various reasons. 
To begin with, it is remarkable that the Camassa-Holm equation 
(\ref{ch}) approximates unidirectional fluid flow in 
Euler's equations at the next order beyond the KdV equation, 
and yet preserves the property of being integrable, fitting as 
it does into the hereditary symmetry framework of Fokas and Fuchssteiner \cite{ff}, 
with a bi-Hamiltonian structure and 
a Lax pair based on a linear spectral problem of second order. 
Also, while there are smooth soliton solutions of (\ref{ch}) on a non-zero 
constant background (or equivalently, with the addition of 
linear dispersion terms), the Camassa-Holm equation has 
peakon solutions, which are peaked solitons of the form 
\beq \label{peaks} 
u(x,t)=\sum_{j=1}^N p_j(t) \, \exp (-|x-q_j(t)|),
\eeq 
where the positions $q_j$ and amplitudes $p_j$ satisfy the 
system of ODEs 
\beq \label{chpeaks} 
\dot{q}_j=\sum_{k=1}^N p_k \, e^{-|q_j-q_k|}, 
\quad \dot{p}_j=p_j\sum_{k=1}^N p_k \,\mathrm{sgn}(q_j-q_k)\, e^{-|q_j-q_k|} 
\eeq 
for $j=1,\ldots , N$. 

The peakons are smooth solutions of (\ref{ch}) 
except at the peak positions $x=q_j$, where the derivative of $u$ is 
discontinuous.  The equations (\ref{chpeaks}) form an 
integrable finite-dimensional Hamiltonian system, corresponding 
to geodesic flow on an $N$-dimensional manifold with inverse 
metric $g^{jk}=\exp(-|q_j-q_k|)$. The positions $q_j$ and momenta $p_j$ 
satisfy the canonical Poisson bracket $\{ q_j , p_k \} = \delta_{jk}$. 
The dynamics of two peakons ($N=2$) 
was solved explicitly in 
the original paper by Camassa and Holm \cite{ch}, 
while the explicit solution for arbitrary $N$ was found by 
Beals, Sattinger and Szmigielski \cite{bss}.
Fuchssteiner also showed that the equation (\ref{ch}) is related via 
a reciprocal transformation to 
the first negative flow in the hierarchy of the Korteweg--de Vries 
equation.

One might wonder whether the Camassa-Holm equation is the only integrable 
PDE of its kind, being a shallow water equation whose dispersionless 
version has weak soliton solutions. This turns out not to be the case. Degasperis 
and Procesi used an asymptotic integrability approach to isolate 
integrable third 
order equations, and discovered a new equation with the dispersionless form  
\beq \label{dp} 
m_t+um_x+3u_xm=0, \qquad m=u-u_{xx}. 
\eeq 
The Degasperis-Procesi equation turns out to be integrable, with a 
bi-Hamiltonian structure and a Lax pair based on a third order spectral 
problem \cite{dhh}, and it also arises in shallow water theory \cite{dgh}. 
The equation (\ref{dp}) is related by a reciprocal transformation 
to a negative flow in the hierarchy of the Kaup-Kupershmidt equation \cite{dhh, wanghone}, 
and it also has peakon 
solutions of the form (\ref{peaks}) 
whose dynamics is described by an integrable finite-dimensional 
Hamiltonian system with a non-canonical Poisson bracket (see \cite{hh}, and section 4 
below). 
The explicit solution of the $N$-peakon dynamics was derived by Lundmark 
and Szmigielski \cite{ls}. 
There are at least two distinct integrable analogues of the Camassa-Holm equation in  
2+1 dimensions \cite{hone,kz}, while the Euler-Poincar\'e  equation on  
the diffeomorphism group (EPDiff) provides a geometrical generalisation of the Camassa-Holm 
equation in arbitrary dimension \cite{epdiff}, and admits weak solutions 
with support on lower-dimensional submanifolds. Rosenau also found 
various PDEs with nonlinear dispersion, which have solutions with 
compact support \cite{rosenau}, some of which are relatives of 
the Camassa-Holm equation \cite{olverli}.

In what follows we present a bi-Hamiltonian structure for 
the integrable hierarchy of PDEs of which (\ref{vn}) is a member, 
present a matrix Lax pair corresponding to a 
zero curvature representation for this equation, 
and show how it is related via a reciprocal transformation to 
a negative flow in the Sawada-Kotera hierarchy. We also 
present a system of Hamiltonian ODEs for the dynamics of 
peakon solutions of (\ref{vn}), and explicitly integrate
the equations for the interaction of two peakons. Finally, we 
compare our results with analogous properties of the 
integrable PDE 
\beq\label{qiao} 
m_t + \Big( m(u^2-u_x^2) \Big)_x=0, \qquad m=u-u_{xx}, 
\eeq 
which was recently obtained by Zhijun Qiao \cite{qiao}.  
Qiao's equation was the original starting point for our study, 
since it has cubic (rather than quadratic) nonlinear terms, and this is 
what led us to ask Vladimir Novikov to seek 
other integrable equations of this kind.

\section{Lax pair and reciprocal transformation} 

The equation (\ref{vn}) arises as a zero curvature 
equation ${\bf F}_t-{\bf G}_x+ [ {\bf F},{\bf G}]=0$, this being the 
compatibility condition for the linear system
\beq \label{laxp}
\begin{array}{ccc} 
\bPsi_x 
& = & {\bf F} {\bPsi},\\ 
                   {\bPsi}_t & = & {\bf G} {\bPsi},
\end{array} 
\eeq 
where 
\beq \label{fg} 
{\bf F}= \left( \begin{array}{ccc} 
0 & m \la & 1 \\ 
0 & 0 & m\la \\ 
1 & 0 & 0 
\end{array} \right), 
\quad
{\bf G}= \left( \begin{array}{ccc} 
\frac{1}{3\la^2}-uu_x & \frac{u_x}{\la}-u^2m\la & u_x^2 \\ 
\frac{u}{\la} & -\frac{2}{3\la^2} & -\frac{u_x}{\la}-u^2m\la  \\ 
-u^2 & \frac{u}{\la} & \frac{1}{3\la^2}+uu_x
\end{array} \right)   .
\eeq 
We found the linear system (\ref{laxp}) directly by applying the 
prolongation algebra method of 
Wahlquist and Estabrook (see \cite{we}, and also \cite{fordy}), but 
the details of this derivation will not be given here\footnote{V. Novikov 
told us that he earlier found a scalar Lax pair for the PDE (\ref{vn}) based on a 
third order spectral problem, by applying a reciprocal transformation to a symmetry of fifth order. 
Any scalar linear problem for (\ref{vn}) 
should be equivalent to the matrix system (\ref{laxp}), possibly after a gauge 
transformation.}. In any case, once a Lax pair is given one can use it to derive most of the important properties 
of an integrable PDE. 

The first important observation we wish to make about Vladimir Novikov's equation is that it is connected to a 
negative flow in the Sawada-Kotera hierarchy via a reciprocal transformation. Upon rewriting   
the PDE (\ref{vn}) in the form 
\beq \label{cons} 
(m^{2/3})_t + (m^{2/3}u^2)_x=0, 
\eeq 
it is immediately clear that $m^{2/3}$ is a conserved density. Since each of the equations (\ref{ch}) 
and (\ref{dp}) has a conserved density of the form 
$m^{1/b}$, for $b=2,3$ respectively, and these densities yield reciprocal transformations to negative flows in more familiar hierarchies,
this suggests that we should define the new independent variables $X$ and $T$ by 
\beq \label{rt} 
dX = m^{2/3} \, dx- m^{2/3}u^2\, dt, \qquad dT=dt. 
\eeq 
The closure condition $d^2 X=0$ for the exact one-form $dX$ in the reciprocal transformation (\ref{rt}) 
is just the conservation law (\ref{cons}). Transforming the time evolution of $m$ in (\ref{vn}), together with the definition 
$m=u-u_{xx}$, leads to the equations 
\beq\label{skminus} 
\Big(\frac{1}{V}\Big)_T=  \Big(\frac{W^2}{V}\Big)_X, \qquad W_{XX}-\left(\frac{V_{XX}}{2V}-\frac{(V_{X})^2}{4V^2}+\frac{1}{V^2}\right)W+1=0, 
\eeq 
where $V=m^{2/3}$ and $W=um^{1/3}$. The evolution equation for $1/V$ in the new independent variables 
$X,T$ is the reciprocal transformation of the equation (\ref{vn}). However, in order to recognise (\ref{skminus}) 
as a member of the Sawada-Kotera hierarchy we need to apply the reciprocal transformation to the Lax pair. (For 
details of the Sawada-Kotera hierarchy and its extensions we refer the reader to \cite{gp}.)
 
By writing the column vector ${\bPsi}$ in components as ${\bPsi}=(\psi_1,\psi_2,\psi_3)^T$, we can eliminate $\psi_1$ and 
$\psi_3$ from ${\bPsi}_x = {\bf F} {\bPsi}$ 
to get a single scalar equation for $\psi=\psi_2$, namely  
\beq\label{scal}
\psi_{xxx} -2m_x m^{-1}\psi_{xx}-(m_{xx} m^{-1}-2(m_x)^2 m^{-2}+1)\psi_x=m^2 \la^2 \psi.
\eeq 
When the reciprocal transformation (\ref{rt}) is used to transform the $x$ derivatives as 
$\partial_x=V\partial_X$,  
the equation (\ref{scal}) becomes 
\beq\label{scalrtX}
\psi_{XXX}+U\psi_X=\lambda^2 \psi, \qquad \mathrm{with}\quad U=-\frac{V_{XX}}{2V}+\frac{(V_{X})^2}{4V^2}-\frac{1}{V^2},
\eeq 
so that the second equation in (\ref{skminus}) has the form $W_{XX}+UW+1=0$ for the same potential $U$. 
The third order operator $\partial_X^3+U\partial_X$ in (\ref{scalrtX})  is the standard Lax operator for the Sawada-Kotera 
hierarchy, and by transforming the $t$ derivatives in ${\bPsi}_t = {\bf G} {\bPsi}$ 
according to $\partial_t=\partial_T-W^2\partial_X$ we find that the $T$ evolution of $\psi$ is given 
by 
\beq\label{scalrtT}
\psi_{T}=\frac{1}{\la^2}\Big(W\psi_{XX}-W_X \psi_X\Big)-\frac{2}{3\la^2}\, \psi. 
\eeq 
After gauging $\psi$ by a factor of $e^{2T/(3\la^2)}$ to remove the final term above, and then replacing $\la^2$ by 
$\la$ and setting $\phi = -3W$, we see that (\ref{scalrtX}) and (\ref{scalrtT}) are respectively equivalent to 
equations (2.25) and (2.26) in \cite{wanghone}, and the compatibility 
requirement $\psi_{TXXX}= \psi_{XXXT}$ 
for this pair of scalar equations gives two conditions, namely 
that $W_{XX}+UW$ is independent of $X$, and $U_T+3W_X=0$. The latter two 
conditions follow from (\ref{skminus}) provided that $U$ is given 
in terms of $V$ as in (\ref{scalrtX}).

\section{Conserved densities and bi-Hamiltonian structure}

The Lax pair (\ref{laxp}) can be used to find infinitely many conserved 
densities for (\ref{vn}). Upon setting $\rho =  (\log \psi )_x$ in 
(\ref{scal}) it is clear that $\rho$ satisfies the equation 
\beq\label{rho} 
\rho_{xx}+3\rho\rho_x + \rho^3-2m_xm^{-1}(\rho_x+\rho^2) 
+\Big(m(m^{-1})_{xx}-1\Big)\rho=m^2\la^2. 
\eeq 
The corresponding $t$ evolution of $\psi$ implies that $\rho_t=F_x$ 
for some flux $F$, and so by expanding $\rho$ in powers of $\la$ 
one finds coefficients that are conserved densities. 
The asymptotic  expansion for $\la\to\infty$ has $\rho^3\sim m^2\la^2 $, so 
$\rho\sim m^{2/3}\la^{2/3}$, which extends to an 
infinite series 
$\rho\sim m^{2/3}\la^{2/3}+\sum_{j=1}^\infty\mu_j\la^{-2j/3} $. 
The densities $\mu_j$ are all determined recursively from 
(\ref{rho}) as local functions of $m$; for example 
$\mu_1 = m^{-5/3}m_{xx}-\frac{4}{3}m^{-8/3}m_x^2+3m^{-2/3}$.   
An expansion in positive powers of $\la$ 
for $\la\to 0$ can consistently begin with 
$\rho\sim - mu \la^2$, but one must solve a second order differential 
equation to obtain each subsequent term, which leads to increasingly 
nonlocal expressions in $m$ and $u$. Since we know that (\ref{vn}) 
is reciprocally related to a negative Sawada-Kotera flow, it is 
natural to regard the $\mu_j$ as densities for Hamiltonians that generate 
a positive hierarchy of flows, with the expansion around $\la =0$
producing Hamiltonian densities for negative flows.   

Having found these conserved densities, we require 
a pair of Hamiltonian operators $\B_1$, $\B_2$ which 
are compatible (in the sense that $\B_1+\B_2$, or any 
linear combination of them, is Hamiltonian) and 
can be used to generate the hierarchy of flows that 
commute with (\ref{vn}). From earlier studies on the 
Camassa-Holm and Degasperis-Procesi equations \cite{wanghone,hh}, 
we know that all nonlocal operators of the form  
\beq\label{bham}
\B =  
m^{1-1/b}D_xm^{1/b}\hat{G}m^{1/b}D_xm^{1-1/b} 
,
\eeq 
with $\hat{G}=(c_1D_x+c_2D_x^3)^{-1}$ for constants 
$b,c_1,c_2$, are Hamiltonian, and have Casimir $\int m^{1/b}dx$. 
In fact, the case $b=2$ gives the third Hamiltonian 
structure for the Camassa-Holm equation, and $b=3$ gives 
the second Hamiltonian structure for the Degasperis-Procesi 
equation. Since $\int m^{2/3}dx$ is a conserved 
quantity for (\ref{vn}), this suggests we should 
consider the operator (\ref{bham}) with $b=3/2$, and indeed 
we find that the equation can be written in Hamiltonian 
form as 
\beq\label{hamform} 
m_t = \B_1 \frac{\de \tilde{H}}{\de m}, \qquad 
\tilde{H}=\frac{1}{4}\int mu\, dx,     
\eeq 
for $\B_1 = -18\B|_{b=3/2}$ 
in the case $c_1=4$, $c_2=-1$. 
Some other conserved quantities are 
$H_1=\int \frac{1}{8}\left(u^4+2u^2u_x^2-\frac{u_x^4}{3}\right)dx$, 
$H_5=\int m^{2/3}dx$, 
$H_7 = \int \frac{1}{3}(m^{-8/3}m_x^2+9m^{-2/3})dx=\int\mu_1dx$, 
and the next one has leading term $H_{11}=\int (m^{-16/3}m_{xxx}^2+\ldots )dx$ 
(up to rescaling).
These are the first 
few in the sequence of Hamiltonians that generate 
local symmetries of weight $k\equiv \pm 1 \bmod 6$ according 
to 
\beq \label{flows} 
m_{t_k}=\B_2\frac{\de H_k}{\de m}=\B_1 \frac{\de H_{k+6}}{\de m},   
\eeq 
where $\B_2 = (1-D_x^2)m^{-1}D_xm^{-1}(1-D_x^2)
$. 
The recursion operator is $\ring = \B_2\B_1^{-1}$, and it generates 
the flows $\ring^n m_x$ of weight $6n+1$ and the flows  
$\ring^n m_{t_5}$ of weight $6n+5$. However, when $k=5$ or $7$  
the rightmost part of the identity (\ref{flows}) fails, 
since both $H_5$ and $H_7$ are Casimirs for $\B_1$; and the 
Hamiltonian $\tilde{H}$ is a Casimir for $\B_2$.    
The proof of the following theorem will be presented 
in a forthcoming article. 

\begin{Thm}\label{hamops}
The operators 
$\B_1 = -2(3mD_x+2m_x)(4D_x-D_x^3)^{-1}(3mD_x+m_x)$ and   
$\B_2 = (1-D_x^2)m^{-1}D_xm^{-1}(1-D_x^2)
$ provide a bi-Hamiltonian structure for the hierarchy of 
symmetries of the equation (\ref{vn}).  
\end{Thm} 
  
\section{Peakon solutions} 

From (\ref{cons}) the travelling waves $u=u(z)$, $z=x-ct$ of (\ref{vn}) satisfy
$(u^2-c)m^{2/3}=\mathrm{const}$. In the general case this gives 
$m=\frac{1}{2}c^2D(u^2-c)^{-3/2}$ for constant $D\neq 0$, which 
integrates further to $(u')^2=u^2+cDu(u^2-c)^{-1/2}+cE$, for another constant $E$. 
This can be reduced to a quadrature which is the sum of elliptic integrals of the third kind, namely 
\beq\label{elliptic3rd}  
dz =\frac{(\frac{1}{w-1}-\frac{1}{w+1})dw}{2\sqrt{(Dw+E)(w^2-1)+w^2}}, 
\quad w=u(u^2-c)^{-1/2}. 
\eeq 
However, if we require waves that vanish at spatial infinity, then $D=0$, which implies 
that $m=0$ whenever $u^2\neq c$. No smooth solution can satisfy the latter 
requirement, but this observation suggests that there should be a weak solution of the form 
\beq\label{onepeak} 
u(x,t)=\pm \sqrt{c}\, e^{-|x-ct-x_0|}, \quad c>0, \quad x_0 \,\, \mathrm{constant}, 
\eeq 
which has the same form as the peakon for the Camassa-Holm  and 
Degasperis-Procesi equations, except that the amplitude is the square root of the speed 
rather than being equal to the speed, as is the case for the peakon solutions of (\ref{ch}) and (\ref{dp}).    
The expression (\ref{onepeak}) has $m=0$ away from the peak, and $u^2=c$ at the peak, but to 
regard it as a weak solution of (\ref{vn}) it is necessary to substitute it into the equation 
and integrate against suitable test functions with support around the peak. For the single peakon 
(\ref{onepeak}) we have $m=\pm 2\sqrt{c}\,\de ( x-ct-x_0 )$, but there is some subtlety in 
interpreting this as a solution, because $u_x=\mp\sqrt{c}\,
\mathrm{sgn}(x-ct-x_0 )e^{-|x-ct-x_0|}$ and $m$ are distributions, 
while 
the equation (\ref{vn}) includes the product $u_x m$. The integrals can be regularised 
by taking the convention $\sgn (0)=0$, but a more rigorous alternative is to 
construct the peakon distribution as a limit of smooth solutions of the PDE. For the Camassa-Holm 
equation it is known that the single peakon arises as a weak solution in this way (see \cite{olverli} for a very detailed treatment), and 
multi-peakons arise similarly as a degenerate limit of algebro-geometric solutions \cite{billiards}.

If we take $u$ to be a linear superposition of $N$ peakons, as in 
(\ref{peaks}), so that $m=2\sum_{j=1}^N p_j(t)\de (x-q_j(t))$, then substituting into the equation (\ref{vn}) 
and integrating against test functions supported at $x=q_j$ gives the equations of motion for the peak positions 
and amplitudes.
\begin{Pro} 
The equation (\ref{vn}) has peakon solutions of the form (\ref{peaks}), whose positions $q_j(t)$ and amplitudes $p_j(t)$ evolve 
according to the dynamical system  
\beq \label{vnpeaks}
\begin{array}{ccl} 
\dot{q}_j & = & \sum_{k,\ell =1}^N p_k p_\ell \, e^{-|q_j-q_k|-|q_j-q_\ell |}, 
\\ 
\dot{p}_j & = & p_j\sum_{k,\ell =1}^N p_k p_\ell \,\mathrm{sgn}(q_j-q_k)\, e^{-|q_j-q_k|-|q_j-q_\ell |} .
\end{array} 
\eeq 
\end{Pro} 
The above equations are not in canonical Hamiltonian form. However, in \cite{hh} one of us showed 
how Hamiltonian operators of the form (\ref{bham}) are reduced to Poisson structures on the 
finite-dimensional submanifold of $N$ peaks or pulses, resulting in the Poisson bracket 
\beq\label{pulsonbracket}
\begin{array}{rcl}
\{ q_j,q_k \} = G(q_j-q_k), & \,\, & \{ q_j,p_k \} = (b-1) G'(q_j-q_k)p_k, \\ 
\{ p_j,p_k \}  & = & -(b-1)^2 G''(q_j-q_k)p_jp_k,
\end{array} 
\eeq 
where $G$ is the skew-symmetric Green's function for the operator $\hat{G}$. For $N>2$, 
the Jacobi identity holds for this bracket if and only if $G$ satisfies the functional equation
\beq\label{func}
G'( 
\al  
)(G( 
\be  
)+G(\ga ))+\mathrm{cyclic}=0 \quad \mathrm{for} \quad \al +\be +\ga =0. 
\eeq 
This functional equation is also a sufficient condition 
for the operator (\ref{bham}) to be Hamiltonian, and Braden and Byatt-Smith proved in the appendix to \cite{hh} 
that the unique continuously differentiable, 
odd solution of equation 
(\ref{func}) is $G(x)=A\,\mathrm{sgn}(x)(1-e^{-B|x|})$  for arbitrary constants $A,B$.  
Up to rescaling $x$, this is the Green's function for the operator $\hat{G}=(D_x-D_x^3)^{-1}$ (or 
$\hat{G}=D_x^{-1}$ in the degenerate case $B\to\infty$). In the case at hand, the operator 
$\B$ in Theorem 
\ref{hamops} 
has $\hat{G}=(4D_x-D_x^3)^{-1}$, 
and the Hamiltonian 
$\tilde{H}$ reduces to a  conserved quantity $h$ for the equations of motion (\ref{vnpeaks}), which 
is quadratic in the amplitudes $p_j$.
\begin{Thm}\label{hamvf} 
The equations (\ref{vnpeaks}) for the motion of $N$ peakons in the PDE (\ref{vn}) are an 
Hamiltonian vector field  
$$ \dot{q}_j=\{ q_j,h \}, \qquad \dot{p}_j=\{ p_j,h \} $$ 
for the Hamiltonian $h=\frac{1}{2}\sum_{j,k=1}^Np_jp_k\exp(-|q_j-q_k|)$, with the Poisson bracket 
specified by 
\beq\label{vnpbs}
\begin{array}{rcl}
\{ q_j,q_k \} = \mathrm{sgn}(q_j-q_k)(1-e^{-2|q_j-q_k|}), & \,\, & \{ q_j,p_k \} = e^{-2|q_j-q_k|} p_k, \\ 
\{ p_j,p_k \}  & = & \mathrm{sgn}(q_j-q_k)e^{-2|q_j-q_k|}p_jp_k.
\end{array}
\eeq  
\end{Thm}    

We conjecture that the equations (\ref{vnpeaks}) constitute a Liouville integrable Hamiltonian system with 
$N$ degrees of freedom.  For $N=1$ this is trivial, and for $N=2$ the result follows from the existence 
of a second independent integral in involution with $h$, namely 
\beq\label{quartic}
k=p_1^2p_2^2(1-e^{-2|q_1-q_2|}), \qquad \{ k,h \} =0. 
\eeq 
The invariant $k$ is degree four in the amplitudes, and for all $N$ there is an analogous integral, quartic in $p_j$, 
obtained by restricting the Hamiltonian $H_1$ to the peakon submanifold. Indeed, the conserved densities for 
the negative flows in the hierarchy of the PDE (\ref{vn}) should all reduce to integrals for the $N$-peakon
dynamics, but the explicit construction of 
$N$ independent Poisson-commuting integrals for (\ref{vnpeaks}) is still in progress. It is also worth mentioning 
that the Lax pair (\ref{laxp}) can be used to obtain an $N\times N$ Lax matrix for the finite-dimensional 
system, satisfying 
\beq \label{laxmatrix} 
\lax\bPhi =-\la^{-2}\bPhi, \qquad \lax = {\bf S} {\bf P}{\bf E}{\bf P}, 
\eeq 
where    ${\bf S}_{jk} = \mathrm{sgn}(q_j-q_k)$, 
${\bf P}=\mathrm{diag}(p_1,\ldots,p_N)$, ${\bf E}_{jk}=\exp (-|q_j-q_k|)$. 
The $j$th component of the vector $\bPhi$ is just 
$\psi_2(q_j(t),t)$, where $\psi_2(x,t)$ is 
the second component of $\bPsi$ in (\ref{laxp}), 
and the corresponding time evolution $\dot{\bPhi}=\mma \bPhi$ yields the Lax equation $\dot{\lax}=[\mma ,\lax ]$ 
for the system (\ref{vnpeaks}). However, unfortunately the spectral invariants of $\lax$, which 
are the coefficients of the characteristic polynomial 
$\mathrm{det} (\lax + \la^{-2} {\bf I})$ (a polynomial in $\la^{-2}$),
do not provide enough integrals. For instance, when 
$N=2$ we find that the trace of $\lax$ vanishes, while the trace of $\lax^2$ gives $k$, but $h$ does not appear. For higher values of $N$ 
we have found that 
the spectral invariants of $\lax$ have degrees $4,8,12,\ldots$ but the integrals of degrees $2,6,10,\ldots$ are missing. This 
leads us to expect that there should be another Lax representation for this system which would provide the correct number of integrals 
for Liouville's theorem.

For the two-peakon dynamics, the equations of motion are 
\beq \label{N2} 
\begin{array}{ccl}
\dot{q}_1 & = & 
(p_1+p_2e^{-|q_1-q_2|})^2 \quad \dot{q}_2=(p_2+p_1e^{-|q_1-q_2|})^2 \\
\dot{p}_1 & = & \mathrm{sgn}(q_1-q_2)e^{-|q_1-q_2|} 
(p_1+p_2e^{-|q_1-q_2|})p_1p_2, \\  
\dot{p}_2 & = & 
-\mathrm{sgn}(q_1-q_2)e^{-|q_1-q_2|}(p_2+p_1e^{-|q_1-q_2|})p_1p_2,  
\end{array}
\eeq 
and without loss of generality we consider the case where the peaks are initially well separated, 
so that $q_1<<q_2$ with $q_1\sim c_1 t$, $q_2\sim c_2 t$  (for $c_1>c_2>0$), and we 
assume that both amplitudes are positive, so $p_1\to\sqrt{c_1}$ and $p_2\to\sqrt{c_2}$ as $t\to -\infty$. 
In terms of these asymptotic speeds the Hamiltonian is $h=\frac{1}{2}(c_1+c_2)$ and the quartic invariant 
is $k=c_1c_2$. Upon integrating the equations (\ref{N2}) we find elementary formulae for $p_2^2-p_1^2$, 
$p_1p_2$ and $e^{-|q_1-q_2|}$, leading to the expressions 
\beq\label{integr}
\begin{array}{cc} 
p_2^2-p_1^2=(c_1-c_2)\tanh T \quad & 
p_1p_2=\sqrt{c_1c_2 +\frac{(c_1-c_2)^4}{16(c_1+c_2)^2} 
\mathrm{sech}^4 T }
\\
q_2-q_1=\frac{1}{2}\log \left(1+\frac{16c_1c_2(c_1+c_2)^2}{(c_1-c_2)^4} 
\cosh^4 T \right), \\ 
\end{array}
\eeq 
where $T=(c_1-c_2)(t-t_0)/2$, with $t_0$ being an arbitrary constant.
The formula for $q_1+q_2$ is somewhat more formidable, being given in terms of 
a certain quadrature as
\beq \label{qsum} 
q_1+q_2= (c_1+c_2)(t-t_0)+\int f(T) dT + \mathrm{const}. 
\eeq
The integrand $f$ is 
$$ 
f(T)= \frac{2(c_1^2-c_2^2)\Big((c_1-c_2)^2+8c_1c_2 \cosh^2 T\Big)}{(c_1-c_2)^4+16c_1c_2(c_1+c_2)^2\cosh^4 T},
$$ 
and the quadrature can be performed explicitly by partial fractions in 
$\tanh (T)$, but the answer is omitted here. 
From (\ref{integr}) and (\ref{qsum}) it is apparent that the peakons exchange speeds under the interaction, without a
head-on collision,
so that $q_1\sim c_2 t$, $q_2\sim c_1 t$  as $t\to \infty$. They also undergo a phase shift, which is 
described by the asymptotics of the term $\int f(T) dT$ in (\ref{qsum}), but the precise formula 
is rather unwieldy and will be presented elsewhere. 

\section{Qiao's equation} 

As we already mentioned, our interest in peakon equations with cubic nonlinearity began with 
Qiao's equation (\ref{qiao}), which can also be written as 
\beq \label{qiao2} 
m_t + (u^2-u_x^2)m_x +2u_xm^2=0. 
\eeq 
Qiao presented a $2\times 2$  Lax pair for this equation given by 
the linear system ${\bPsi}_x={\bf U} {\bPsi}$, ${\bPsi}_t  =  {\bf V} {\bPsi}$ with 
\beq \label{qlax} 
\begin{array}{ccl} 
{\bf U} & = &  \left( \begin{array}{cc} 
-\frac{1}{2}& \frac{1}{2}m \la \\ 
-\frac{1}{2}m \la & \frac{1}{2}  
\end{array} \right), 
\\
{\bf V} &= & \left( \begin{array}{ccc} 
\la^{-2}+  \frac{1}{2}(u^2-u_x^2)&-\la^{-1}(u-u_x)-\frac{1}{2}m\la (u^2-u_x^2)\\ 
 \la^{-1}(u+u_x)+\frac{1}{2}m\la (u^2-u_x^2) & 
-\la^{-2}-  \frac{1}{2}(u^2-u_x^2)
\end{array} \right)   .
\end{array}
\eeq 
Qiao also found a bi-Hamiltonian structure for his equation, namely 
\beq 
\label{qbham} 
m_t= \tilde{\B}_1\frac{\de \tilde{H}}{\de m}= \tilde{\B}_2\frac{\de H_{1}}{\de m}
\eeq 
where 
\beq \label{qops} 
\tilde{\B}_1= -4D_x m D_x^{-1} m D_x, \quad \tilde{\B}_2=-2(D_x-D_x^3),  
\eeq 
and 
$\tilde{H}$, $H_1$ are the same as the conserved quantities for (\ref{vn}) 
given in section 3 above. (In Qiao's original papers the quantity $H_0$, proportional 
to $\tilde{H}$ here, is out by a factor of 2, while the quantity denoted $H_1$ in
\cite{qiao} is missing the $u_x^4$ term.) Note that the first 
operator in (\ref{qops}) is of the form (\ref{bham}) with $b=1$, and the compatibility 
of these Hamiltonian structures can be proved by a slight extension 
of a result in \cite{wanghone}. 

If we apply the reciprocal transformation 
$$
dX=\frac{m}{2}dx-\frac{1}{2}m(u^2-u_x^2)dt, \quad dT=dt
$$ to Qiao's equation (\ref{qiao}) then we find the pair of equations 
\beq\label{qrec}
(m^{-2})_T=-2u_X, \quad (mu_X)_X=4(u/m-1).
\eeq 
By transforming the Lax pair given by (\ref{qlax}) and writing a scalar 
linear problem for $\psi_1$, the first component  of $\bPsi$, we
find that the $X$ part is
$$
\psi_{1,XX}+(v_X-v^2)\psi_1=-\la^{2}\psi, \quad v=m^{-1},
$$  
which is the Schr\"odinger equation corresponding to the 
spectral problem for KdV, and the expression $v_X-v^2$ 
is the standard Miura map from modified KdV. The corresponding time evolution is 
$$ 
\psi_{1,T}=-\frac{1}{\la^2}\Big(a\psi_{1,X}-\frac{1}{2}a_x\psi_1\Big), 
\quad a =u-mu_X/2, 
$$ from which it is clear that the pair of equations (\ref{qrec}) 
corresponds to a negative flow in the (modified) KdV hierarchy. 

Qiao has noted that the equation (\ref{qiao}) does not have standard peakons 
of the form $u=ce^{-|x-ct|}$. The general travelling wave 
solution for this equation can be solved in terms of an elliptic 
integral, and some interesting  wave shapes have been found in \cite{qiao} in 
cases where this integral reduces to expressions in hyperbolic functions. 
However, here we should like to point out that, at least formally, peakons of the form  
$u=\pm \sqrt{c}e^{-|x-ct|}$ (just as found for (\ref{vn}) above) do provide solutions
of Qiao's equation. 
From the  equation in the form (\ref{qiao2}) it is clear that if $m$ is given by a
delta function then the $m^2$ terms do not make sense. However, if we 
take travelling waves $u=u(z)$, $z=x-ct$ and integrate (\ref{qiao}) 
along the $z$ axis against an arbitrary test function $\varphi$, and then perform 
an integration by parts, we find 
\beq \label{parts} 
\int m\Big(u^2-(u')^2-c\Big)\varphi '(z) \, dz =0.
\eeq   
For the peakon $u(z)=\sqrt{c}\, e^{-|z|}$ we have 
$u'(z)=-\sqrt{c}\,\mathrm{sgn}(z) e^{-|z|}$ and $m(z)=2\sqrt{c}\,\de (z)$, 
and this satisfies (\ref{parts}) as long as we assume the usual convention 
that $\mathrm{sgn}(0)=0$. A more careful derivation could be 
carried out along the lines of \cite{olverli}. The equations for 
$N$ peakons should be extremely degenerate, since $b=1$ and 
$G(x)$ is proportional to $\mathrm{sgn}(x)$ in the bracket (\ref{pulsonbracket}), so 
$p_j$ are constant 
and the amplitudes of the peakons do not change. The same 
conclusion is reached by integrating  (\ref{qiao}) against a test function and 
performing integration by parts.

It seems that peakon equations with cubic nonlinearity have several novel 
features compared with the Camassa-Holm and Degasperis-Procesi equations, 
and there are many more things to be revealed by further study. 

\vspace{.1in} 

\noindent {\bf Acknowledgements.} We are very grateful to Vladimir 
Novikov for sharing his latest classification results with us. AH thanks Darryl 
Holm for suggesting integrating by parts with Qiao's equation, and is 
very grateful to  Alexander Strohmaier for his remarks about 
distributional solutions.

\end{document}